\documentclass[12pt,a4paper]{article}
\usepackage[dvips]{graphicx,color}
\newcommand{\be}{\begin{equation}}
\newcommand{\ee}{\end{equation}}
\newcommand{\ba}{\begin{eqnarray}}
\newcommand{\ea}{\end{eqnarray}}
\newcommand{\baa}{\begin{eqnarray*}}
\newcommand{\eaa}{\end{eqnarray*}}

\begin{document}

{\pagestyle{empty}
\vskip 2.5cm

{\renewcommand{\thefootnote}{\fnsymbol{footnote}}
\centerline{\large \bf
Prediction of transmembrane helix configurations}
\vskip 0.5 cm
\centerline{\large \bf
by replica-exchange simulations}
}
\vskip 2.0cm

\centerline{Hironori Kokubo$^{a,}$\footnote{\ \ e-mail: kokubo@ims.ac.jp}
and Yuko Okamoto$^{a,b,}$\footnote{\ \ e-mail: okamotoy@ims.ac.jp}
}
\vskip 1.5cm
\centerline{$^a${\it Department of Functional Molecular Science}}
\centerline{{\it The Graduate University for Advanced Studies}}
\centerline{{\it Okazaki, Aichi 444-8585, Japan}}
\centerline{and}
\centerline{$^b${\it Department of Theoretical Studies}}
\centerline{{\it Institute for Molecular Science}}
\centerline{{\it Okazaki, Aichi 444-8585, Japan}}

\medbreak
\vskip 2.0cm

\centerline{\bf Abstract}
\vskip 0.3cm

We propose a method for predicting helical membrane protein structures by computer simulations. 
Our method consists of two parts. In the first part, amino-acid sequences of the transmembrane 
helix regions are obtained from one of existing WWW servers. In the second part, we perform 
a replica-exchange simulation of these transmembrane helices with some constraints and identify 
the predicted structure as the global-minimum-energy state. 
We have tested the method with the dimeric transmembrane domain of glycophorin A. 
The structure obtained from the prediction was in close agreement with the experimental data.

}
\section{Introduction}
\label{introduction}
It is one of the most important problems to predict protein tertiary 
structures from the amino-acid sequence information in the structural genomics era.  
It is estimated that 20-30
\% of all genes in most genomes encode membrane proteins \cite{TMHMM,genome2}. However, 
only a small number of detailed structures have been obtained for membrane proteins 
because of technical difficulties in experiments such as high quality crystal growth. 
Therefore, it is desirable to develop a method for predicting membrane protein
structures by computer simulations (for previous attempts, see, for instance, Refs.
\cite{pre-simu1}--\cite{pre-simu4}). 
In particular, structures of membrane proteins 
are simpler than those of soluble proteins and most membrane proteins are composed of 
several transmembrane helices (we do not consider $\beta$-sheet membrane proteins in 
the present work).  Taking it into consideration, it is possible to search more 
limited conformational space than that of usual simulations. In other words, 
the prediction of membrane protein structures may be easier than that of soluble proteins 
because there is only one type of secondary structures. 

In the two-stage model \cite{two1}, individual helices of a membrane protein are postulated 
to be stable separately as domains in a lipid bilayer and then side-to-side helix association 
is driven, resulting in a functional protein. In fact, some experimental evidence indicates 
that the formation of $\alpha$-helices and the positioning of transmembrane helices are 
independent \cite{two2,two5}. Separated fragments of bacteriorhodopsin formed 
independently $\alpha$-helical conformations in the membrane, the native structure could 
be recovered mixing the fragments. Therefore it is reasonable to assume that processes of 
helix formation and positioning may be predicted separately.

Our prediction method thus consists of two parts. 
In the first part, amino-acid sequences of the transmembrane 
helix regions are obtained from database analyses
\cite{TMHMM}\cite{database1}--\cite{HMMTOP}. 
In the second part, we perform 
a replica-exchange simulation of these transmembrane helices with some constraints and identify 
the predicted structure as the global-minimum-energy state. 
However, it is difficult to obtain a 
global-minimum state in potential energy surface by conventional molecular dynamics (MD) or 
Monte Carlo (MC) simulations. This is because there exist a huge number of 
local-minimum-energy states, and the simulations tend to get trapped in one of the 
local-minimum states. One popular way to overcome this multiple-minima problem is to 
perform a generalized-ensemble simulation (for reviews, see 
Refs.~\cite{gene-review1,gene-review2}), which is based on non-Boltzmann probability 
weight factors so that a random walk in potential energy space may be realized. The random walk 
allows the simulation to go over any energy barrier and sample much wider configurational 
space than by conventional methods. One of well-known generalized-ensemble algorithms
is the replica-exchange 
method (REM) \cite{rep-exch1}--\cite{rep-exch2} (the method is also referred to as parallel 
tempering \cite{para-temp}). We apply this method to the structure prediction of 
membrane proteins. The dimeric transmembrane domain of glycophorin A is often used as 
a model system of helix-helix interaction of membrane proteins \cite{pre1,pre2}. 
In this Letter, we test our prediction method with this system.

Sec.~2 summarizes the details of our method for predicting transmembrane helix 
configurations. In Sec.~3 the results of the application to the structure prediction 
of the dimeric transmembrane domain of glycophorin A are given. Section 4 is devoted 
to conclusions.

\section{Methods}
\label{methods}
Our method consists of two parts. In the first part, amino-acid sequences of the transmembrane 
helix regions of the target protein are identified. It is already established that 
the transmembrane helical 
segments can be predicted by analyzing mainly the hydrophobicity of amino acid sequences, 
without having any information about the higher order structures. 
There exist many WWW servers such as
TMHMM \cite{TMHMM}, 
MEMSAT \cite{MEMSAT}, 
SOSUI \cite{SOSUI}, and
HMMTOP \cite{HMMTOP}
in which given the amino-acid sequence 
of a protein they judge 
whether the protein is a membrane protein or not and (if yes) predict the regions in 
the amino-acid sequence that correspond to the transmembrane helices.  

In the second part, we perform a REM simulation of these transmembrane 
helices that were identified in the first part. Given the amino-acid sequences of 
transmembrane helices, we first construct ideal canonical $\alpha$-helices (3.6 residues 
per turn) of these sequences. For our simulations, we introduce the following
rather drastic approximations: 
(1) We treat the backbone of the $\alpha$-helices as rigid body and only 
side-chain structures are made flexible. (2) We neglect the rest of the amino acids of 
the membrane protein (such as loop regions). (3) We neglect surrounding molecules
such as lipids. In  principle, 
we can also use molecular dynamics method, but we employ Monte Carlo algorithm here. 
We update configurations with rigid translations and rigid rotations of each $\alpha$-helix 
and torsion rotations of side chains. We use a standard force field such as 
CHARMM \cite{cha1,cha2} for the potential energy of the system. We also add the 
following simple harmonic constraints to the original force-field energy:

\begin{eqnarray}
E_{\rm constr} &=& \sum_{i=1}^{N_{\rm H}-1} k_1~ \theta \left( r_{i,i+1}-d_{i,i+1} \right)
\left[ r_{i,i+1}-d_{i,i+1} \right]^2 \nonumber \\
&+& \sum_{i=1}^{N_{\rm H}} \left\{ k_2~ \theta \left( \left| z^{\rm L}_{i}-z^{\rm L}_{0} 
\right| -d^{\rm L}_{i} \right)
\left[ \left| z^{\rm L}_{i}-z^{\rm L}_{0} \right| -d^{\rm L}_{i}\right]^2\right. \nonumber \\
&+& ~~~~~\left. k_2~ \theta \left( \left| z^{\rm U}_{i}-z^{\rm U}_{0} \right| -d^{\rm U}_{i} \right)
\left[ \left| z^{\rm U}_{i}-z^{\rm U}_{0} \right| -d^{\rm U}_{i} \right]^2 \right\} \nonumber \\
&+& \sum_{{\rm C}_{\alpha}} k_3~ \theta \left( r_{{\rm C}_{\alpha}}-d_{{\rm C}_{\alpha}} \right) 
\left[ r_{{\rm C}_{\alpha}}-d_{{\rm C}_{\alpha}} \right]^2,
\label{const-ene}
\end{eqnarray}
where $N_{\rm H}$ is the total number of transmembrane helices in the protein and 
$\theta(x)$ is the step function:
\begin{eqnarray}
\theta(x)=\left\{ \begin{array}{ll}
1 & {\rm for} ~x \geq 0, \\
0 & {\rm otherwise}, \\
\end{array} \right.
\label{step-func}
\end{eqnarray} 
and
$k_1$, $k_2$, and $k_3$ are the force constants of the harmonic constraints, 
$r_{i,i+1}$ is the distance between the C atom of the C-terminus of the $i$-th 
helix and the N atom of the N-terminus of the $(i+1)$-th helix, $z^{\rm L}_{i}$ and 
$z^{\rm U}_{i}$ are the z-coordinate values of the C$_{\alpha}$ (or C) atom of the 
N-terminus (or C-terminus) of the $i$-th helix near the fixed lower boundary 
value $z^{\rm L}_0$ and the upper 
boundary value $z^{\rm U}_0$ of the membrane, respectively, $r_{{\rm C}_{\alpha}}$ are the 
distance of C$_{\alpha}$ atoms from the origin, and $d_{i,i+1}$, $d^{\rm L}_{i}$, 
$d^{\rm U}_{i}$, and $d_{{\rm C}_{\alpha}}$ are the corresponding central value constants
of the harmonic constraints.

The first term in Eq.~(\ref{const-ene}) is the energy that constrains pairs of 
adjacent helices along the amino-acid chain not to be apart from each other 
too much (loop constraints). This term has a non-zero value only when the 
distance $r_{i,i+1}$ becomes longer than $d_{i,i+1}$. 

The second term in Eq.~(\ref{const-ene}) is the energy that constrains helix N-teminus 
and C-terminus to be located near membrane boundary planes. This term has a non-zero value 
only when the C atom of each helix C-terminus and C$_{\alpha}$ atom of each 
helix N-terminus are apart more than $d^{\rm L}_{i}$ (or $d^{\rm U}_{i}$). Base on the 
knowledge that most membrane proteins are placed in parallel, this constraint energy 
is included so that helices are not too apart from the perpendicular orientation
with respect to the membrane boundary planes.

The third term in Eq.~(\ref{const-ene}) is the energy that constrains all C$_{\alpha}$ 
atoms within the sphere (centered at the origin) of radius $d_{{\rm C}_{\alpha}}$. 
This term has a non-zero value only when C$_{\alpha}$ atoms go out of this sphere. 
This term is introduced so that the center of mass of the molecule stays near 
the origin. The radius of sphere is set to a large value in order to guarantee that
a wide configurational space is sampled.

We now briefly 
review the replica-exchange method (REM) \cite{rep-exch1}--\cite{rep-exch2}
(see Refs.~\cite{rep-exch3,gene-review2} for details). 
The system for REM consists 
of $M$ non-interacting copies (or, replicas) of the original system in the canonical 
ensemble at $M$ different temperatures $T_{m}$ $(m=1,...,M)$. Let $X=(...,x_{m}^{[i]},...)$ 
stand for a state in this ensemble. Here, the superscript $i$ and the 
subscript $m$ in $x_{m}^{[i]}$ label the replica and the temperature, respectively. 
The state $X$ is specified by $M$ sets of $x_{m}^{[i]}$, which in turn is
specified by the coordinates $q^{[i]}$ of all the atoms in replica $i$.
A simulation of the REM is then realized by alternately performing the following two steps. 
Step~1: Each replica in canonical ensemble of the fixed temperature is simulated 
simultaneously and independently for a certain MC or MD steps. Step~2: A pair of replicas, 
say $i$ and $j$, which are at neighboring temperatures $T_{m}$ and $T_{n}$, respectively 
are exchanged: 
$X = (...,x_{m}^{[i]},...,x_{n}^{[j]},...) \rightarrow X' = (...,x_{m}^{[j]},...,x_{n}^{[i]},...)$. 
The transition probability of this replica exchange is given by the Metropolis criterion:
\begin{eqnarray}
w(X \rightarrow X') = w(x_{m}^{[i]}|x_{n}^{[j]})=\left\{ \begin{array}{ll}
1 & {\rm for} ~\Delta \leq 0, \\
\exp(-\Delta) & {\rm otherwise,} \\
\end{array} \right.
\label{tran-prob}
\end{eqnarray} 
where
\begin{equation}
\Delta = \left( \beta_m-\beta_n \right) \left( E(q^{[j]})-E(q^{[i]}) \right)~.
\label{delta}
\end{equation}
Here, $E(q^{[i]})$ and  $E(q^{[j]})$ are the potential energy of the $i$-th and the $j$-th 
replica, respectively. In the present work, we employ Monte Carlo algorithm for Step~1.
There are $2 N_{\rm H} + N_{\rm D}$ kinds of Monte Carlo moves, where $N_{\rm D}$
is the total number of dihedral angles in the side chains of $N_{\rm H}$ helices.
The first term corresponds to the rigid translation and rigid rotation of the 
helices and the second to the dihedral-angle rotations in the side chains.  One MC
sweep is defined to consist of $2 N_{\rm H} + N_{\rm D}$ updates that are randomly chosen
from these MC moves with the Metropolis evaluation for each update. 

We predict the native structure of membrane spanning regions from the 
global-minimum-energy state obtained by the REM simulations. 

\section{Results and discussion}
\label{results}

In the first part of the present method, we obtain amino-acid sequences of the 
transmembrane helix regions from existing WWW servers such as
those in Refs.~\cite{TMHMM,MEMSAT,SOSUI,HMMTOP}.
However, the precision of these programs in the WWW servers is about 85 \% and 
needs improvement. We thus focus our attention on the effectiveness of the second 
part of out method, leaving this improvement to the developers of the WWW servers. 
Namely, we use the experimentally known amino-acid sequence of 
helices and try to predict their conformations, following the prescription of 
the second part of our method described in the previous section. We selected the 
amino-acid sequence of the transmembrane dimer of glycophorin A (PDB code: 1AFO). 
The number of amino acids for each helix is 18 and the sequence is TLIIFGVMAGVIGTILLI. 

At first, the ideal canonical $\alpha$-helix (3.6 residues per turn) of this 
sequence was constructed. The N and C termini of this helix were blocked with 
acetyl and N-methyl groups, respectively. The force field that we used is the 
CHARMM param19 parameter set (united-atom model) \cite{cha1,cha2}. 
No cutoff was introduced to the non-bonded energy terms, and the dielectric 
constant $\epsilon$ was set equal to 1.0. We have also studied the case of 
$\epsilon=4.0$, because it is the value close to that for the lipid environment.
The computer code based on the CHARMM macromolecular mechanics program \cite{cha3} 
was used and the replica-exchange method was implemented in it. 

This helix structure was minimized subject to harmonic restraints on all the heavy atoms. 
The initial configuration for the REM simulation was that two $\alpha$-helices
of identical sequence and structure thus prepared
were placed in parallel at a distance of 20 \AA.  These helices  
are quite apart from each other and the starting configuration is indeed very 
different from the native one.  Note that the only information derived from the 
NMR experiments \cite{nmr1} is the amino-acid sequence of the individual helices.

\begin{figure}
\begin{center}
\includegraphics[width=16cm,keepaspectratio]{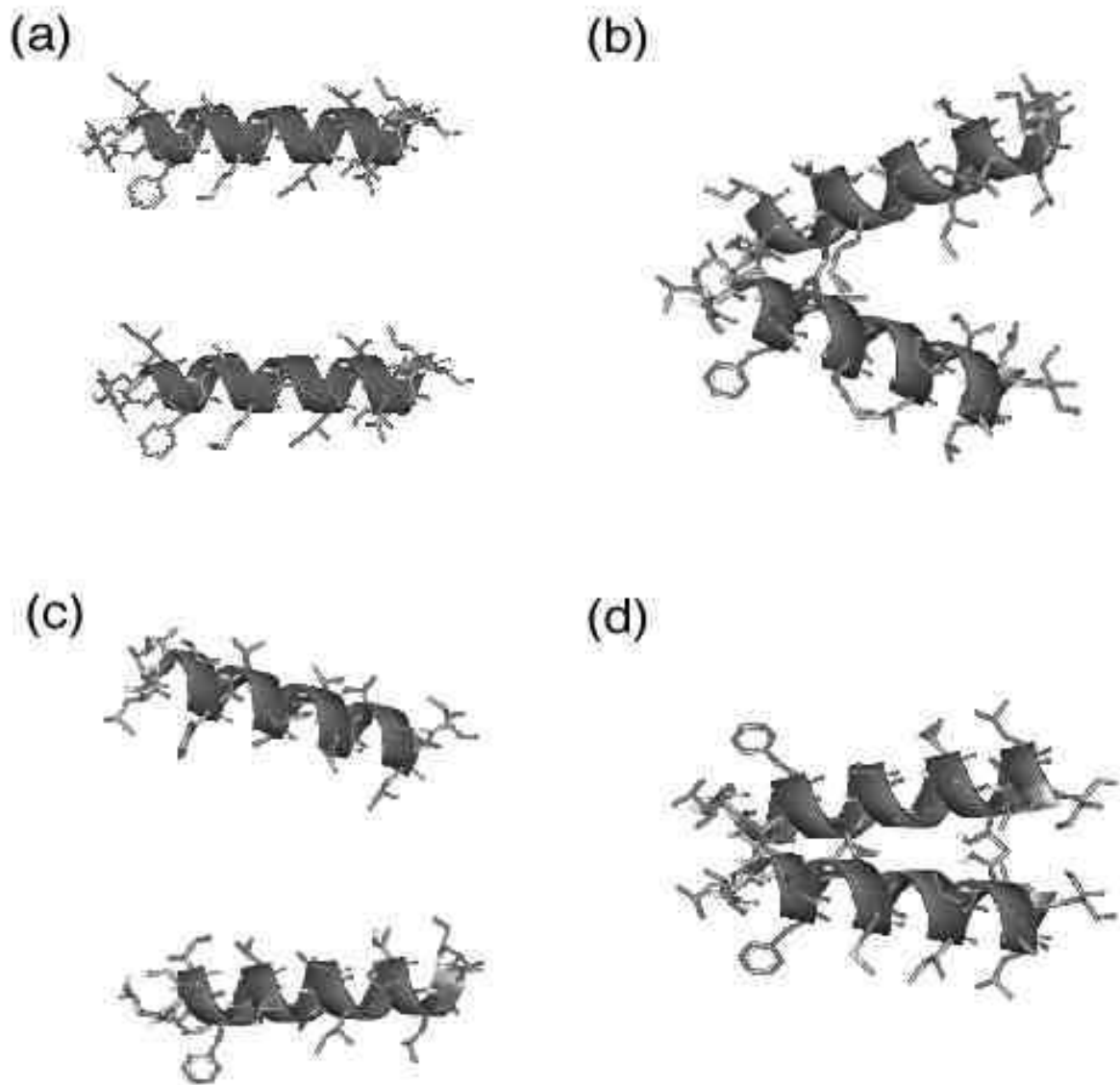}
\caption{Typical snapshots from the REM simulation of Fig. 2. The initial
configuration (a), the configurations at 294,931 MC sweep (b), at 732,000 MC sweep (c), 
and at 810,000 MC sweep (d).}
\label{fig_3}
\end{center}
\end{figure}

The values of the constants in the constraints in Eq.~(\ref{const-ene}) were set
as follows: $N_{\rm H} = 2$, $k_1=k_2= 0.5$ kcal/(mol \AA$^2$), $k_3=0.05$ kcal/(mol \AA$^2$),
$d_{i,i+1} = 20$ \AA, $z^{\rm L}_0 = -13.35$ \AA, $z^{\rm U}_0 = +13.35$ \AA,
$d^{\rm L}_i = d^{\rm U}_i = 1.0$ \AA, and $d_{{\rm C}_{\alpha}} = 50$ \AA.
The values for $z^{\rm L}_0$ and $z^{\rm U}_0$ were taken from the z-coordinates
of the initial configuration (in Fig.~\ref{fig_3}(a) below; the z axis is placed
horizontally in the figure).
The values of $d^{\rm L}_i$ and $d^{\rm U}_i$ are set loosely in the sense that
the constraint energy will not be turned on until the interhelix angle of 
two helices becomes more than about $45^{\circ}$.
In the present example of 
glycophorin A dimer, the first term in Eq.~(\ref{const-ene}) were imposed on
both terminal ends (i.e., two kinds of $r_{i,i+1}$ were prepared: one is
the distance between a pair of N atoms at the N-terminus of the two helices
and the other is the distance between a pair of C atoms at the C-terminus
of the two helices).

We performed two REM MC simulations of 1,000,000 MC sweeps, starting from this 
parallel configuration: one with the dielectric constant $\epsilon = 1.0$
and the other with $\epsilon = 4.0$.  We used the following 13 temperatures: 
200, 239, 286, 342, 404, 489, 585, 700, 853, 1041, 1270, 1548, and 1888 K, 
which are distributed almost exponentially. The highest temperature was chosen 
sufficiently high so that no trapping in local-minimum-energy states occurs. 
This temperature distribution was chosen so that all the acceptance ratios 
are almost uniform and sufficiently large ($>$ 10 \%) for computational efficiency. 
Backbone structures were fixed during simulations and Monte Carlo move type was 
taken to be rigid translation of each helix, rigid rotation of each helix, 
and torsion-angle rotations of side chains.

\begin{figure}
\begin{center}
\includegraphics[width=16cm,keepaspectratio]{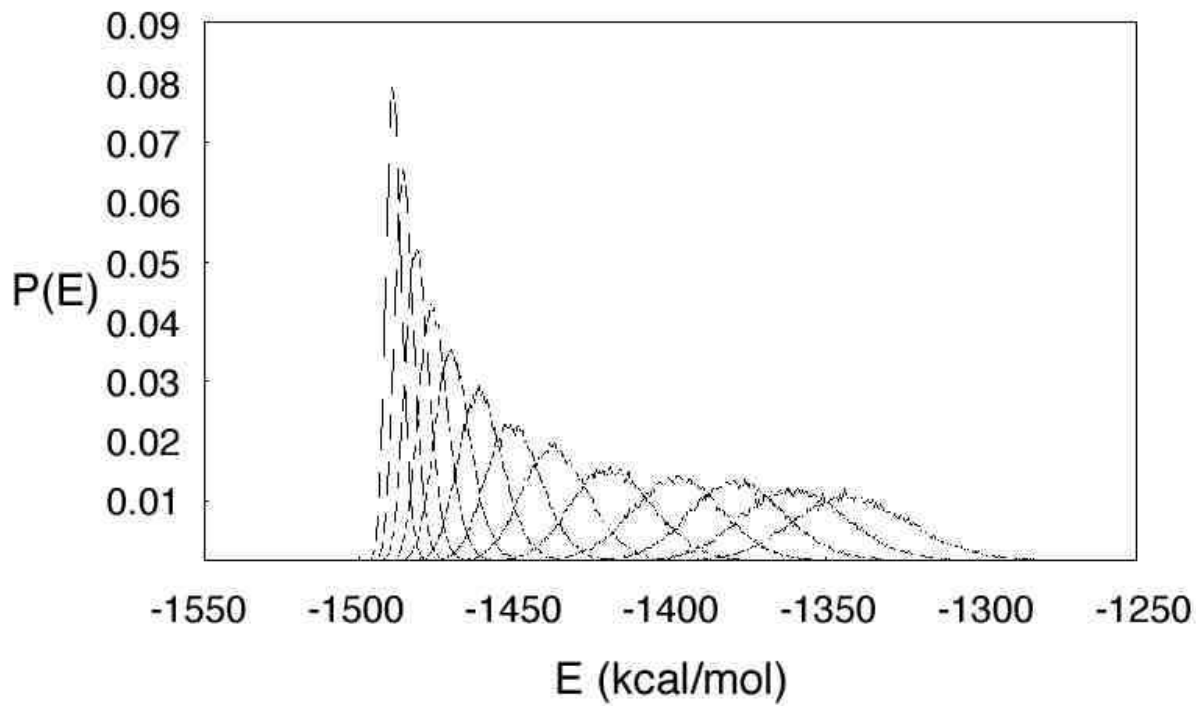}
\caption{The canonical probability distributions of the total potential energy obtained 
from the replica-exchange MC simulation at the thirteen temperatures.
The distributions correspond to the following temperatures (from left to right):
200, 239, 286, 342, 404, 489, 585, 700, 853, 1041, 1270, 1548, and 1888 K,}
\label{fig_1}
\end{center}
\end{figure}

We first discuss the results with $\epsilon = 1.0$.
In Fig.~\ref{fig_1} the canonical probability distributions of the total potential
energy obtained at the 
chosen 13 temperatures from the REM simulation are shown. We see 
that there are enough overlaps between all neighboring pairs of distributions, 
indicating that there will be sufficient numbers of replica exchange between 
pairs of replicas.
The obtained acceptance ratio of the replica exchange are listed in Table~\ref{acceptance-ratio}. 
We see that the acceptance ratio of replica exchange between all pairs of 
neighboring temperatures is large enough as expected in Fig~\ref{fig_1}. 
The results in Table~\ref{acceptance-ratio} 
imply that one should observe a free random walk in the replica and temperature space.

\begin{figure}
\begin{center}
\includegraphics[width=16cm,keepaspectratio]{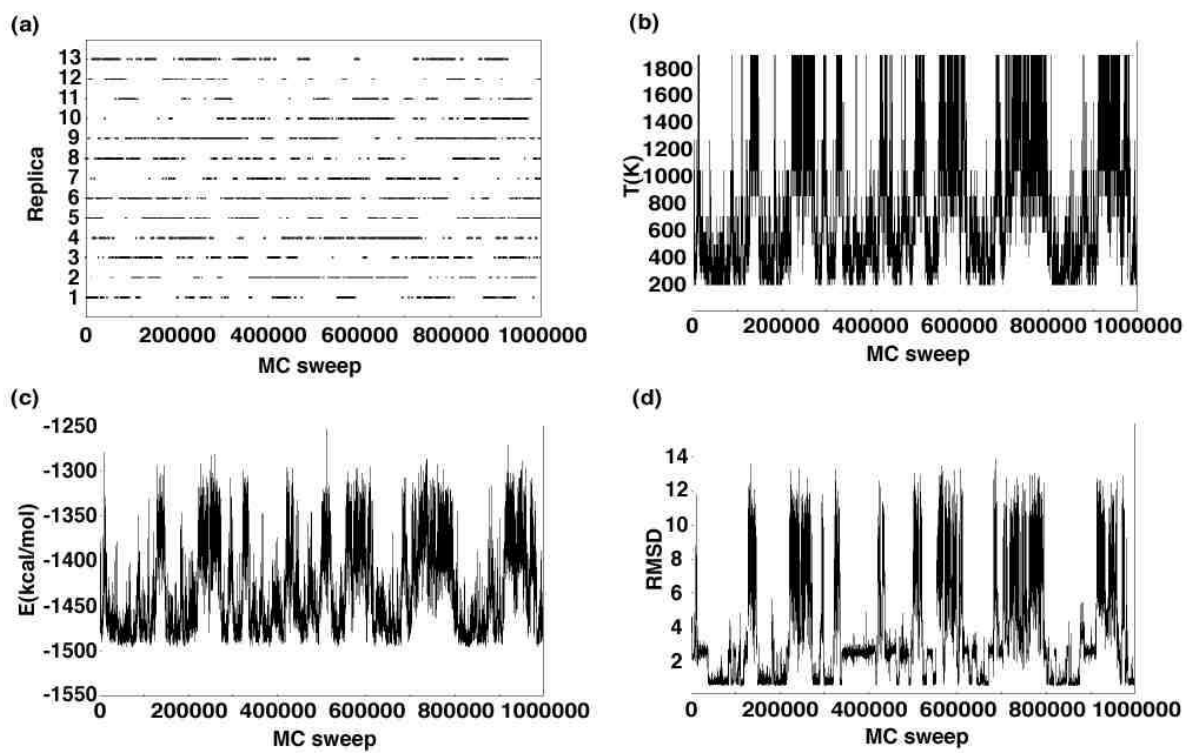}
\caption{Time series of replica exchange at $T = 200$ K (a), temperature exchange for one of 
the replicas (Replica 8) (b), the total potential energy for Replica 8 (c), and the RMS
deviation (in \AA) of backbone atoms from the NMR structure for Replica 8 (d).}
\label{fig_2}
\end{center}
\end{figure}

In Fig.~\ref{fig_2}(a) we show the ``time series'' of replica exchange at the lowest temperature
($T = 200$ K). We see that every replica takes the lowest temperature many times, and we indeed 
observe a random walk in the replica space. The complementary picture to this is the temperature 
exchange for each replica. The results for one of the replicas (Replica 8) are shown in 
Fig.~\ref{fig_2}(b). We again observe a random walk in the temperature space between the 
lowest and highest temperatures. Other replicas perform random walks in the same way. 
In Fig.~\ref{fig_2}(c) the corresponding time series of the total potential energy is shown. 
We see that a random walk in the potential energy space between low and high energies is 
also realized. Note that there is a strong correlation between the behaviors in 
Figs.~\ref{fig_1}(b) and \ref{fig_1}(c) as there should.
All these results confirm that the present REM simulation has been properly performed.

We now study how widely the configurational space is sampled during the present simulation. 
For this purpose, we plot the time series of the root-mean-square (RMS) deviation of the 
backbone atoms from the NMR structure \cite{nmr1} in Fig.~\ref{fig_2}(d). When the 
temperature becomes high, the RMS deviation takes a large value (the largest
value in Fig.~\ref{fig_2}(d) is 13.9 \AA, and the maximum value among all the
replicas is 15.7 \AA), and when the temperature 
becomes low, the RMS deviation takes a small value (the smallest value in
Fig.~\ref{fig_2}(d) is 0.48 \AA, and the minimum value among all the
replicas is 0.47 \AA).  By comparing Fig.~\ref{fig_2}(c) and Fig.~\ref{fig_2}(d),
we see that there is a strong correlation between the total potential
energy and the RMS deviation values.  In particular, it is remarkable that
when the energy is the lowest (around $-1490$ kcal/mol), most of the RMS
values are as small as about 0.5 \AA.  This implies that the global-minimum-energy
state is indeed very close to the native structure.

In Fig.~\ref{fig_3} typical snapshots from the REM simulation of 
Fig.~\ref{fig_2} are shown. Fig.~\ref{fig_3}(a) is the initial configuration
of this simulation, in which the two helices are in parallel.
We see that this simulation sampled many 
non-native configurations such as those in Fig.~\ref{fig_3}(b) and Fig.~\ref{fig_3}(c).
At low temperatures low-energy configurations such as that in
Fig.~\ref{fig_3}(d) with the side chains packed are sampled. 
We see that the REM simulation performs a random walk not only
in energy space but also in 
conformational space and that it does not get trapped in one of a huge number of 
local-minimum-energy states. 
  
\begin{figure}
\begin{center}
\includegraphics[width=16cm,keepaspectratio]{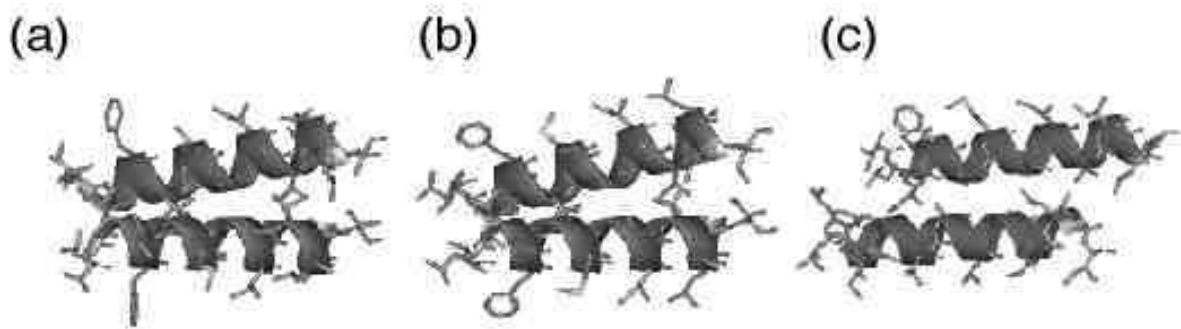}
\caption{(a) The NMR configuration (PDB code 1AFO, Model 16). (b) The global-minimum-energy 
configuration
predicted by the REM simulation with the dielectric constant $\epsilon$ = 1.0. 
(c) The global-minimum-energy configuration predicted by the REM simulation with $\epsilon = 4.0$. 
The RMS deviation from the native configuration of (a) is 0.59 \AA~(b) and 4.48 \AA~(c) 
with respect to all backbone atoms, and it is 1.31 \AA~(b) and 5.55 \AA~(c) with respect
to all atoms.}
\label{fig_4}
\end{center}
\end{figure}

In Fig.~\ref{fig_4} the configuration obtained by the NMR experiments \cite{nmr1}
and the global-minimum-energy configurations
obtained by the REM simulations are compared.
Here, we also show the result with the dielectric constant $\epsilon = 4.0$.
At first sight, it is rather surprising that the result with $\epsilon = 1.0$
is much closer to the experimental result than that with $\epsilon = 4.0$,
because the dielectric constant for a lipid system is closer to 4.0 than to 1.0.
However, on second thoughts we understand that the present results are
reasonable because the pairs of helices in transmembrane proteins are tightly 
packed and almost no lipid molecules can exist between helices. 
This implies that
helix-helix interactions are the main driving force in the final stage of
the structure formation of membrane proteins.

\section{Conclusions}
\label{conclusions}
In this Letter we proposed a method for predicting the membrane spanning structures of 
helical membrane proteins.
Our method consists of two parts.  In the first part, amino-acid sequences of the
transmembrane helix regions of the target protein are obtained from one of
existing WWW servers.  The precision of these programs in the WWW servers is
at present about 85 \%, but it is expected to be further improved.
In the second part of our method, we perform a generalized-ensemble
simulation of these transmembrane helices with atomistic details to obtain
the global-minimum (free) energy state, which we identify as the predicted
structure.  In order to save computation time, we introduced rather bold
approximations: Backbones are treated as rigid body (only side-chain structures
are made flexible) and the rest of the protein
such as loop regions and the surrounding lipids are neglected.
With these assumptions, however, the example that we tested gave a remarkable 
agreement of the predicted structure with the experimental data.
We believe that the inclusion of atomistic details of side chains was particularly
important because transmembrane helices are usually tightly packed.  This fact
also justifies the validity of our assumptions in the sense that almost
no lipid molecules can exist between helices.
In the future we have to make our approximations better.
For instance, we should introduce some flexibility in the helix backbone structures.
The electrostatic interactions, in which we used the dielectric constant value of
1.0, can also be modified so that some environmental effects may be
taken into account.

\noindent
{\bf Acknowledgements}

While the present work was in final stage, we learned that W. Im, M. Feig,
and C.L. Brooks, III of the Scripps Research Institute, U.S.A. have
obtained similar results on glycophorin A by replica-exchange MD simulations.
We thank C.L. Brooks, III for showing their results prior to publication.
The computations were performed on the computers at the Research Center for
Computational Science, Okazaki National Research Institutes and at ITBL,
Japan Atomic Energy Research Institute.
This work was supported, in part, by
the NAREGI Nanoscience Project, Ministry of Education, Culture, Sports, Science
and Technology, Japan.



\begin{table}
\caption{Acceptance ratio of replica exchange corresponding to pairs of neighboring temperatures}
\label{acceptance-ratio}
\begin{center}
\begin{tabular}{lccc} \hline
Pairs of temperatures & Acceptance ratio \\ \hline
 200 $\longleftrightarrow$  239 K         & 0.41             \\
 239 $\longleftrightarrow$  286 K         & 0.40             \\
 286 $\longleftrightarrow$  342 K         & 0.39             \\
 342 $\longleftrightarrow$  404 K         & 0.40             \\
 404 $\longleftrightarrow$  489 K         & 0.32             \\
 489 $\longleftrightarrow$  585 K         & 0.34             \\
 585 $\longleftrightarrow$  700 K         & 0.33             \\
 700 $\longleftrightarrow$  853 K         & 0.28             \\
 853 $\longleftrightarrow$ 1041 K         & 0.29             \\
1041 $\longleftrightarrow$ 1270 K         & 0.36             \\
1270 $\longleftrightarrow$ 1548 K         & 0.42             \\
1548 $\longleftrightarrow$ 1888 K         & 0.46             \\ \hline
\end{tabular}
\end{center}
\end{table}


\end{document}